\documentclass[12pt]{article}
\usepackage{amsmath, amssymb, slashed, epsf, color, graphicx}

\def\beq{\begin{eqnarray}}
\def\eeq{\end{eqnarray}}
\def\bea{\begin{eqnarray}}
\def\eea{\end{eqnarray}}

\def\stilde{\widetilde}

\newcommand{\gsim}{\lower.7ex\hbox{$\;\stackrel{\textstyle>}{\sim}\;$}}
\newcommand{\lsim}{\lower.7ex\hbox{$\;\stackrel{\textstyle<}{\sim}\;$}}
\newcommand{\newc}{\newcommand}
\newc{\Nc}{N_{c}}
\newc{\CG}{C_G}
\newc{\gp}{g'}
\newc{\stopi}{\stilde t_i}
\newc{\sboti}{\stilde b_i}
\newc{\staui}{\stilde \tau_i}
\newc{\stopj}{\stilde t_j}
\newc{\sbotj}{\stilde b_j}
\newc{\stauj}{\stilde \tau_j}
\newc{\stopI}{\stilde t_1}
\newc{\stopII}{\stilde t_2}
\newc{\sbotI}{\stilde b_1}
\newc{\sbotII}{\stilde b_2}
\newc{\stauI}{\stilde \tau_1}
\newc{\stauII}{\stilde \tau_2}
\newc{\sstop}{s_{t}}
\newc{\cstop}{c_{t}}
\newc{\ssbot}{s_{b}}
\newc{\csbot}{c_{b}}
\newc{\sstau}{s_{\tau}}
\newc{\cstau}{c_{\tau}}
\newc{\Sstop}{s_{2t}}
\newc{\Cstop}{c_{2t}}
\newc{\Ssbot}{s_{2b}}
\newc{\Csbot}{c_{2b}}
\newc{\Sstau}{s_{2\tau}}
\newc{\Cstau}{c_{2\tau}}
\newc{\salpha}{s_\alpha}
\newc{\calpha}{c_\alpha}
\newc{\Calpha}{c_{2\alpha}}
\newc{\Salpha}{s_{2\alpha}}
\newc{\sbetapm}{s_{\beta_\pm}}
\newc{\cbetapm}{c_{\beta_\pm}}
\newc{\Sbetapm}{s_{2 \beta_\pm}}
\newc{\Cbetapm}{c_{2 \beta_\pm}}
\newc{\sbetaO}{s_{\beta_0}}
\newc{\cbetaO}{c_{\beta_0}}
\newc{\SbetaO}{s_{2 \beta_0}}
\newc{\CbetaO}{c_{2 \beta_0}}
\newc{\vu}{v_u}
\newc{\vd}{v_d}
\newc{\seL}{\stilde e_L}
\newc{\smuL}{\stilde \mu_L}
\newc{\seR}{\stilde e_R}
\newc{\smuR}{\stilde \mu_R}
\newc{\suL}{\stilde u_L}
\newc{\sdL}{\stilde d_L}
\newc{\suR}{\stilde u_R}
\newc{\sdR}{\stilde d_R}
\newc{\scL}{\stilde c_L}
\newc{\ssL}{\stilde s_L}
\newc{\scR}{\stilde c_R}
\newc{\ssR}{\stilde s_R}
\newc{\snue}{\stilde \nu_e}
\newc{\snumu}{\stilde \nu_\mu}
\newc{\snutau}{\stilde \nu_\tau}
\newc{\Gpm}{G^\pm}
\newc{\Hpm}{H^\pm}
\newc{\FFbS}{\overline{FF}S}
\newc{\FFbV}{\overline{FF}V}
\newc{\FSS}{F_{SS}}
\newc{\FSSS}{F_{SSS}}
\newc{\FFFS}{F_{FFS}}
\newc{\FFFbS}{F_{\overline{FF}S}}
\newc{\FSSV}{F_{SSV}}
\newc{\FVS}{F_{VS}}
\newc{\FVVS}{F_{VVS}}
\newc{\FFFV}{F_{FFV}}
\newc{\FFFbV}{F_{\overline{FF}V}}
\newc{\Fgauge}{F_{\rm gauge}}
\newc{\DRbarprime}{$\overline{\rm DR}'$ }
\newc{\DRbar}{$\overline{\rm DR}$ }
\newc{\MSbar}{$\overline{\rm MS}$ }
\newc{\Yu}{{\bf Y}_u}
\newc{\Yd}{{\bf Y}_d}
\newc{\Ye}{{\bf Y}_e}
\newc{\Au}{{\bf a}_u}
\newc{\Ad}{{\bf a}_d}
\newc{\Ae}{{\bf a}_e}
\newc{\bm}{{\bf m}}
\newc{\zhol}{Z^{\rm hol}}
\newc{\rwino}{r_{\tilde W}}
\newc{\rmu}{r_{\tilde H}}
\newc{\ra}{r_A}
\newc{\ccdot}{\!\cdot\!}

\begin{document}

\begin{titlepage}
\begin{center}

\hfill IPMU-0011 \\
\hfill FTPI-MINN-13/01\\
\hfill UMN-TH-3132/13\\

\vspace{1.0cm}
{\Large \bf Natural SUSY's Last Hope:} \\
\vspace{0.2cm}
{\large \sl $R$-parity Violation via $UDD$ Operators}
\vspace{1.5cm}

{Biplob Bhattacherjee}$^{(a)}$,
{Jason L. Evans}$^{(a, b)}$,
{Masahiro Ibe}$^{(a, c)}$, \\
{Shigeki Matsumoto}$^{(a)}$,
and
{Tsutomu T. Yanagida}$^{(a)}$

\vspace{1.0cm}
{\it
$^{(a)}${Kavli IPMU, University of Tokyo, Kashiwa, 277-8583, Japan} \\
$^{(b)}$William I. Fine Theoretical Physics Institute, \\
School of Physics and Astronomy, \\
University of Minnesota, Minneapolis, MN 55455, USA \\
$^{(c)}${\it ICRR, University of Tokyo, Kashiwa, 277-8583, Japan}
}
\vspace{1.5cm}

\abstract{Here, we give a broad overview of the more natural spectra allowed by the LHC when $UDD$ $R$-parity violation is allowed. Because $R$-parity violation removes the missing energy signals in colliders, the experimental constraints on the gluino, stops, sbottoms and higgsinos are relatively mild. We also show that $UDD$ $R$-parity violation and lepton number conservation can be made consistent with grand unification. This feat is achieved through the product unification, $SU(5)\times U(3)$. In this model, mixing of the SM quarks with additional quark like particles charged under the $U(3)$ generate a $UDD$ $R$-parity violating operator. Furthermore, these models are also capable of generating a ``natural" spectra. The emergence of these more natural low-scale spectra relies heavily on the fact that the gaugino masses are non-universal, a natural consequence of product unification.  }

\end{center}
\end{titlepage}
\setcounter{footnote}{0}

\section{Introduction}
\label{sec: introduction}

Prior to the LHC, naturalness was widely accepted as motivation for studying supersymmetry (SUSY). Experimental constraints on superpartner masses were rather mild. In fact, superpartner masses could still be as light as ${\cal O}(100)$~GeV. After three years of LHC running, most SUSY spectra require the gluino be heavier than 1~TeV. Since the higgsino and stop masses are the important factors in the one-loop potential, naively it seems natural SUSY is unaffected\footnote{Since natural SUSY requires both the left- and right-handed stop masses to be ${\cal O}(100)$~GeV, the left-handed sbottom mass must also be ${\cal O}(100)$~GeV because of the SU(2)$_L$ symmetry.}. However, a Higgs soft mass of ${\cal O}(100)$~GeV at the weak scale is also required. The renormalization group running of Higgs soft mass is drastically affected by the gluino mass at the two loops level. If the gluino mass is larger, it will generate large radiative corrections to the soft mass. These large radiative corrections must be canceled against the Higgs boundary mass to realize a soft mass of ${\cal O}(100)$~GeV at the weak scale. Thus, we have significant tuning in the Higgs sector. As a result, the higgsino, stops, and gluino must all have a masses of ${\cal O}(100)$~GeV to alleviate the tuning in the Higgs sector.

Since natural SUSY requires a light gluino, a significant number of events are inevitably generated through gluino pair production at the LHC. If $R$-parity is conserved, these events usually involve large transverse missing energy $\slashed{E}_T$. Using kinematical selections for $\slashed{E}_T$, standard model backgrounds can be significantly reduced, leading to strong constraints on the SUSY spectra. In order to evade the constraints, several authors have considered variations of the vanilla mSUGRA mass spectra. One example is the compressed SUSY spectra~\cite{LeCompte:2011fh}. This method is able to avoid detection because the visible energy is too small to pass the selections, and events with large $\slashed{E}_T$ are excluded from analysis. Another possibility is stealth SUSY spectra, where additional electroweak scale particles are introduced to alter the decay patters of the MSSM particles~\cite{Stealth}. In both of these models, however, tuning is introduced to realize a natural SUSY spectra\footnote{By model building this spectrum can natural be generated as it was in~\cite{Murayama:2012jh}}. Fortunately, if these model with a very special spectra are correct, they can be tested in near future. For instance, the gluino mass in compressed SUSY spectra has already been constrained to be $\gtrsim 500~(600)$~GeV by observing initial state radiation gluons at 7 (8)~TeV run~\cite{Bhattacherjee:2012mz}. The constraint will be as strong as $\gtrsim 1$~TeV if SUSY is not seen at 14~TeV run. For stealth SUSY spectra, the gluino mass has already been constrained to be $\gtrsim 1$~TeV when an isolated photon is produced in a squark decay chain~\cite{CMS:2012un}.

On the other hand, it is also possible to evade the $\slashed{E}_T$ constraints when we consider $R$-parity violating scenarios. As is well known, if $R$-parity is violated, the lightest supersymmetric particle (LSP) is not stable and signals with large $\slashed{E}_T$ disappear. However, if the violation is generic, experimental constraints can be quite drastic. For example, if both the baryon ($B$) and lepton ($L$) $R$-parity violating couplings are allowed, the lifetime of the proton will be quite short. Proton decay constraints would require the $R$-parity violating couplings to be too small to be relevant in colliders. To evade stringent constraints like these, either $B$ or $L$ must be conserved. If lepton number is violated, the gluino would decay to multiple leptons which can easily be seen. These multi-lepton signals constrain the gluino mass to be larger than 1~TeV. On the other hand, if $R$-parity is only broken by couplings that violate baryon number, the major constraint on the gluino mass comes from multi-jet searches. Depending on the flavor structure of the $R$-parity violation, the constraints can be weaker and allow for a gluino mass much less than 1~TeV while still being consistent with current LHC data. Furthermore, a light gluino may still be allowed even if SUSY is not seen at the early stage of 14 TeV run, because multi-jets QCD backgrounds against the signal are also expected to significantly increase. The baryon $R$-parity violating scenario will therefore be the last hope for natural SUSY. In this letter, we focus on the so-called $UDD$ $R$-parity violating scenario.

Although this form of $R$-parity violation seems enticing, it does require some model building, which is the main topic in this letter. A major complication for this scenario is how to generate these $R$-parity violating couplings in a GUT consistent way without also generating lepton number violation. To solve this conundrum, we will appeal to the product gauge unification $SU(5) \times U(3)$~\cite{ProdUni}. This product unification allows us to hide the $R$-parity violation in quark like triplets that are charged under the $U(3)$. When the $SU(3)$ subgroup of $SU(5)$ breaks diagonally with the $U(3)$, the $R$-parity violation in the $U(3)$ sector is converted to $R$-parity in the $SU(3)_c$ sector of the MSSM. Thus this model allows for $R$-parity violation which breaks $U(1)_B$ while conserving $U(1)_L$. In what follows, we first consider phenomenology of $UDD$ $R$-parity violating scenario and then show that a gluino mass much less than 1~TeV is indeed possible without conflicting with the latest LHC data (section \ref{sec: pheno}). To show examples of these natural SUSY spectra, we will explicitly show that a gluino mass around 600~GeV is still allowed, by performing Monte-Carlo simulations of their LHC signatures. Next, in section \ref{sec: model}, our model of $R$-parity violation is discussed. The last section will contain our conclusions.

\section{Phenomenology of $UDD$ $R$-parity violation}
\label{sec: pheno}
If naturalness is to be taken seriously, the higgsinos, stop quarks, one sbottom squark and the gluino
should be relatively light. However, the other sparticles may be made heavy ($\sim$ a few TeV) without
affecting naturalness. We have already mentioned the severity of the LHC bounds on $R$-parity conserving scenario as well as the $R$-parity
violating scenarios where lepton number is violated. However, if the lightest SUSY particle decays only to
quarks, it is possible to have a more natural SUSY spectrum. For this reason, we only consider only the $UDD$ $R$-parity
violating couplings and take lepton number to be conserved.  Constraints on $UDD$ type operators come from direct collider
searches and additionally from indirect experimental observations like proton decay, $n-\bar{n}$
oscillation, renormalization group evolution etc. In this section, we first discuss indirect constraints
and then study collider constraints on SUSY particles. \\

Let us now focus on the constraints of the couplings
\begin{eqnarray}
W_{R_p} =\frac{\lambda_{ijk}''}{2} \bar U_i \bar D_j \bar D_k,\label{RPC}
\end{eqnarray}
where $\lambda_{ijk}''$ is some coupling that violates $R$-parity and $i,j,k$ are flavor indices. The couplings in Eq. (\ref{RPC}) violate baryon number. If these interactions are combined with the sphaleron, which violates $B+L$, the entire baryon asymmetry can be washed out. However, if the baryon asymmetry is generated after or around the time of the electroweak phase transition, via Affleck-Dine baryogenesis~\cite{AfDineBar} or even possibly electroweak baryogenesis~\cite{EWBG}, we are able to avoid this rather strong constraint on the couplings. Since we wish to consider rather large $R$-parity violating couplings, we will assume the baryon asymmetry was produced via one of these mechanism.

Now, we discuss the most relevant constraints for each of the 9 couplings in Eq. (\ref{RPC}) \cite{Barbier:2004ez}. We begin with the most severely constrained coupling, $\lambda_{11k}''$. These couplings are constrained by $n-\bar n$ oscillations\footnote{Although tuning the left-right mixing to zero will suppress this contribution, we are considering natural models and so we assume this constraint to be valid.}. By adjusting the constraint in \cite{Zwirner:1984is} to fit the mass spectra we are interested in and including the left-right mixing neglected there, we find
\begin{eqnarray}
|\lambda_{11k}''| \lesssim (10^{-5}-10^{-4})\left(\frac{10^8s}{t_{osc}}\right)\left(\frac{m_{\tilde k}}{600 {\rm GeV}}\right)^{4}\left(\frac{m_{\tilde g}}{600 {\rm GeV}}\right)^{1/2}\left(\frac{2500~{\rm GeV}}{m_k X_k}\right)
\label{eq: ntonbar}
\end{eqnarray}
Here $m_{\tilde k}$ are the down-type squark masses and $m_{\tilde g}$ is the gluino.  $X_k$ is the left right mixing term for the down-type squark $k$.  If we assume some tuning in $X_k$, there would be an even weaker bound on these couplings.

For the coupling\footnote{In constraining this coupling we have assumed that the gravitino and any axions are heavier than the proton. Otherwise, the constraints on this coupling are quite severe.} $\lambda_{112}''$, there is a stronger and even more stringent bound which comes from the $NN\to KK$ decays. These decays proceed via a similar graph as $n-\bar n$ oscillations minus the left-right squark mixing suppression. Modifying the constraint found in \cite{Barbier:2004ez,Goity:1994dq} to fit our mass spectrum, we find

\begin{eqnarray}
|\lambda_{112}''| \lesssim 10^{-5} \left(\frac{m_{\tilde g}}{600 {\rm GeV}}\right)^{1/2}\left(\frac{m_{\tilde s}}{600 {\rm GeV}}\right)^2
\label{eq: NNtoKK}
\end{eqnarray}
However, as stated in \cite{Goity:1994dq}, there is a lot of uncertainty in the hadron matrix element. In fact this error can be as large as a few orders of magnitude.

Next, we discuss the constraint on the couplings $\lambda_{312}''$ and $\lambda_{313}''$.  These couplings are also constrained by $n-\bar n$ oscillations involving very complicated diagrams \cite{Barbier:2004ez,Chang:1996sw} and are
\begin{eqnarray}
&& |\lambda_{321}''|\lesssim [2.1\times 10^{-3},1.5\times 10^{-2}]\left(\frac{m_s}{200 {\rm MeV}}\right)^{-2} \\
&& |\lambda_{331}''|\lesssim [2.6\times 10^{-3},2.0\times 10^{-2}]
\end{eqnarray}
where the terms in brackets correspond to $m_{\tilde q}=[100,200]~{\rm GeV}$ and $m_s$ is the strange quark mass. These constraints are much weaker than is need for our later discussion and they will get weaker as the squark mass increases, so, we give no further details of these constraints.

The remaining couplings are constrained by renormalization group running.  This constrains the couplings to be no larger than \cite{Goity:1994dq}

\begin{eqnarray}
|\lambda_{123}''| \lesssim 1.25,\\
|\lambda_{212}''| \lesssim 1.25,\\
|\lambda_{213}''| \lesssim 1.25,\\
|\lambda_{223}''| \lesssim 1.25,\\
|\lambda_{323}''| \lesssim 1.12.
\end{eqnarray}

The presence of $R$-parity violation makes the collider phenomenology much more complicated than the usual $R$-parity
conserving case, because of the various choices for the LSP as well as the presence of new
$R$-parity violating couplings. Unlike the $R$-parity conserving case, the presence of the $\lambda^{''}$ couplings allows the LSP to decay to SM particles and thus the final states crucially depend on the
choice of the LSP and the new couplings. The naturalness arguments demand light higgsinos,
stops, gluino, and one sbottom. However, naturalness does not dictate the relative mass hierarchies of these particles and so allows for many different natural spectra. Let us now consider a scenario for which the gluino is the LSP.
If the gluino LSP decays through the $\lambda^{''}$ R-parity-violating interaction in equation (\ref{RPC}) through a
virtual squark, its decay length becomes longer for larger squark masses
and/or if the $R$-parity violating coupling is smaller. To evade LHC constraints, the gluino decay length must be
small enough to be considered prompt. In some cases, the phenomenological constraints on the $R$-parity violating
couplings are too severe and the gluino decay length is too long. The decay length of an LSP gluino for degenerate squarks
is estimated to be
\begin{eqnarray}
c\tau_{\rm gluino}
=
\frac{256\pi^2}{3 \alpha_s (\lambda^{\prime\prime})^2}
\frac{m_{\rm squark}^4}{m_{\rm gluino}^5}
\simeq
166 \, \mu{\rm m}
\left(\frac{0.001}{\lambda^{\prime\prime}}\right)^2
\left(\frac{m_{\rm squark}}{1 \, {\rm TeV}} \rule{0mm}{3.2ex} \right)^4
\left(\frac{100 \, {\rm GeV}}{m_{\rm gluino}}\right)^5,
\label{eq: decay length}
\end{eqnarray}
where $\lambda^{\prime\prime}$ is ether $\lambda^{\prime\prime}_{112}$, $\lambda^{\prime\prime}_{212}$,
 or $\lambda^{\prime\prime}_{312}$. It can be seen from this equation that the decay length of the gluino LSP becomes much too
 long when the coupling $\lambda^{\prime\prime}$ is highly suppressed. Examining the previous section we see that the coupling $\lambda^{\prime\prime}_{112}$
is highly suppressed and will not be relevant, as we will discuss below. At the LHC, decay lengths larger than
 100 $\mu$m can be detected in principle by examining the distribution of the impact parameters~\cite{Asano:2011ri} of
the gluino decay products. If the decay length is too large, it is also possible to detect the gluino as a displaced
vertex~\cite{Aad:2011zb}. We therefore simply consider the parameter space with a decay length smaller than 100$\mu$m
where the gluino decays can be regarded as a prompt.

By examining the limit on $\lambda^{\prime\prime}_{112}$ found in equation (\ref{eq: NNtoKK}), we see that the
decay length of the gluino is always longer than 100 $\mu$m for a gluino mass lighter than 500 GeV. If we use a more
accurate formula for the gluino decay length than that in Eq. (\ref{eq: decay length}), the bound on the gluino mass
becomes even more severe, $m_{\rm gluino} >$ 800 GeV. As a result, the gluino decays facilitated by the coupling
 $\lambda^{\prime\prime}_{112}$ may not be favorable, making decays through the coupling $\lambda^{\prime\prime}_{212}$ or
$\lambda^{\prime\prime}_{312}$ more interesting. For a stop squark or higgsino LSP, a very small $\lambda^{''}$
coupling is excluded for similar reasons.  \\

Calculation of the bounds on SUSY particles for all possible mass hierarchies is an extremely cumbersome task and this is
not our aim. Rather, we take a few examples motivated by naturalness and discuss the bounds coming from
collider experiments. Throughout our discussion, we will only study the production and decays of the higgsinos, third generation squarks
and gluino assuming other sparticles to be heavy enough. It was pointed out in the previous paragraph that SUSY
particles with long lifetime are more constrained and for this reason we only consider scenarios with prompt decays. Let us now
illustrate possible collider constraints on individual particles relevant for natural SUSY. After which, we will talk about
more complicated possibilities. \\

{\bf Higgsino:}  Because the $\mu$ term appears as the tree level Higgs mass, it should be small
enough to avoid large fine tuning. In other words, naturalness demands light higgsinos.
If the higgsinos are lighter than the other SUSY particles, the lightest higgsino can only decay through $R$-parity violating couplings and the decay products are three quarks (including top and bottom quarks).
Higgsinos may be pair produced and eventually each decays to 3 quarks giving six or more jets in the final
state. The final states may have leptons which are produced from top quark decays. LEP collaborations studied
such scenarios and has placed a limit on $M_2$ and $\mu$ for a  particular value of the unified GUT sfermion
mass and $\tan \beta$ \cite{Heister:2002jc}. Although the limit is model dependent, higgsino masses greater than
100 GeV can be safely taken since it is close to the LEP kinematic reach. Hadron colliders are not very sensitive
to such final states and the higgsino production cross section is relatively small. It is thus very difficult
to constrain higgsinos using LHC data if other sparticles are heavy enough.\\

{\bf Stop and sbottom squarks:} Stop squarks play very important role in determining the Higgs mass through one
loop effects and naturalness requires the stops to be light. Stop squark can directly decay to two quarks through $\lambda^{''}$
couplings or it can decay to other SUSY particles such as higgsinos/gluino/sbottom through $R$-parity conserving channel.
Similarly, sbottom squark can decay to two quarks or to higgsinos/gluino/stops. For large $R$-parity violating coupling,
single production may also play important role. Pair production of stop squark can be directly identified by looking for pairs
of di-jet resonances. The LEP experiments searched for excess in 4 jets final state and put a limit of 82.5
GeV on doublet up type quark \cite{Heister:2002jc} which is also applicable for doublet stop squark.
A recent study \cite{Evans:2012bf} indicates that if stop decays to two light quarks, there is no bound from
LHC 7 TeV data because the QCD multi-jet background overwhelms the signal.
However, if the final state contains b quark, the SM background is under control and 200 GeV stop
squark could be discovered with 20 fb$^{-1}$ of integrated luminosity at the 8 TeV LHC \cite{Franceschini:2012za}.
The above arguments are also true for sbottom squark.\\

{\bf Gluino:} Although the gluino affects the Higgs mass at two loop level, its contribution to RG running is still non negligible
and the presence of a heavy gluino destroys naturalness. If the gluino is the LSP, it can only decay to 3 quarks through
$R$-parity violating couplings. Severe constraints on this production and decay mode come from searches for three-jet
resonances. Currently, the CDF collaboration of Tevatron experiment has excluded the gluino mass below 144 GeV,
 with the assumptions that first and second generation squarks are heavier than 500 GeV~\cite{Aaltonen:2011sg}.
In addition, the CMS collaboration of LHC experiment has recently excluded the region 200 GeV $< m_{\rm gluino} <$
460 GeV~\cite{Chatrchyan:2011cj, :2012gw}. This analysis is very similar to the CDF analysis, i.e, based on three-jet resonance
search. Comparing both bounds, it seems that the gluino mass window between 144 GeV and 200 GeV is still allowed.
However, a recent study by ATLAS collaboration rules out this region. Unlike resonance searches, their analysis
is based on counting the signal and background events and dividing it into two parts, each optimized
separately for a high and low mass gluino. They put very stringent bounds on the gluino mass and a gluino mass between 100 GeV
 and 666 GeV \cite{atlas6j} is ruled out by the 7 TeV data set. The key point of this strategy is to select six high $p_T$
jets above a certain threshold. However, in the case where the gluino
is the NLSP and a neutralino is the LSP, the gluino will decay to a neutralino by emitting two quarks and the decay of
the neutralino will produce three more jets. We have closely followed the analysis performed by ATLAS collaboration. It seems that it is possible to relax the upper bound of gluino mass  up to 100 GeV in that case. \\

In our analysis of $R$-parity-violating decays of the gluino, constraints on the couplings $\lambda^{\prime\prime}_{113}$
and $\lambda^{\prime\prime}_{123}$ are expected to be stronger than those on $\lambda^{\prime\prime}_{112}$, as defined
in Eq. (\ref{RPC}). This is because the gluino decays through the former two couplings always produce $b$ quarks.
SM backgrounds with $b$ quarks can be reduced much more efficiently because of b-tagging. This huge reduction in background
is not expected when the gluino decays through the coupling $\lambda^{\prime\prime}_{112}$. For this same reason,
constraints on the couplings $\lambda^{\prime\prime}_{212}$ is expected to be weaker than those on $\lambda^{\prime\prime}_{213}$,
 $\lambda^{\prime\prime}_{223}$ respectively, although, the analysis of a gluino LSP which decays through b quarks in the context of 7/8 TeV LHC
has not been performed yet. \\

There is another interesting possibility which requires special attention. If the first index of $\lambda^{\prime\prime}$ coupling
is 3 (like $\lambda^{\prime\prime}_{312}$), the gluino can decay to top quarks and we may get leptonic final state in some gluino
decays. Since the gluino is a Majorana particle, we may get same sign top quarks and thus same sign leptons from gluino pair
production. CMS has searched for same sign di-electron or muon events with two or more b-tagged jets and missing transverse energy
using 4.7 fb $^{-1}$ data at 7 TeV \cite{Chatrchyan:2012sa}. They define nine signal regions depending on $\slashed{E}_T$ and $H_T$ cuts. From the non observation
of excess over SM background, they put an upper limit on the signal cross section. We use this analysis to calculate
a gluino mass bound. The signal events are generated using Pythia~\cite{Sjostrand:2006za}, while Delphes~\cite{Ovyn:2009tx}
was used to simulate detector effects. The signal cross section was calculated at NLO level using Prospino~\cite{Beenakker:1996ed}.
We have found that gluino LSP could be as light as 560 GeV assuming $\lambda^{\prime\prime}_{312}$ coupling.
Similar types of bound were also obtained in ref. \cite{Allanach:2012vj}. The ATLAS collaboration
has searched for a gluino decaying to top quarks in the context of $R$-parity conserving scenario using 5.8 fb$^{-1}$ data at 8 TeV
LHC run \cite{ATLAS:2012sna}. They look for final state with same sign $e$ or $\mu$ with $p_T>$ 20 GeV, at least 4 jets with $p_T>$ 50 GeV and $\slashed{E}_T>$150 GeV.
We perform similar kind of analysis and we have found a bound on gluino LSP decaying to top quark in $R$-parity violating
scenario is about 580 GeV. \\

From the above discussion, it is clear that the gluino LSP is more strongly constrained by collider searches. However, it should be noted that
this bound may be different if the gluino is not the LSP. It is also important to know the situation for which the stops/sbottom
and higgsinos are lighter than gluino. Let us assume that higgsino is the LSP and stop/sbottom squarks are placed
in between higgsinos and the gluino. Since we are talking about light gluino ($\sim$ 500-600 GeV), the relative mass difference
between the gluino and higgsinos play very important role in determining the final states. For small mass difference between the gluino
and the stop/sbottom , $\tilde{g} \rightarrow  b \tilde{b}_1 $ will be dominant because $\tilde{g} \rightarrow  t \tilde{t}_1 $
decay mode will be phase space suppressed. In this case, $\tilde{b}_1$ will decay to a top quark and charged higgsino. On the
other hand, if the gluino decays to a stop squark, the stop will decay to b quark and chargino. This means that in both cases we will get
a top quark , a b quark plus three jets from a single gluino decay. Here the jet multiplicity could be as high as 14 and
the number of jets above the threshold $p_T$ should be smaller than in the gluino LSP case. This means that gluino mass bound of
666 GeV should be relaxed a little bit. Here CMS resonance search will not be applicable. However, the presence of top quark
as a decay product can produce same sign leptonic final states, and the gluino mass limit should not be too different
from to the gluino mass bound when it decays to top quark directly. Here, the jet multiplicity is very large and thus
lepton isolation could be problematic. A naive estimation indicates that gluino mass $\sim$ 600 GeV and lighter stop/sbottom and higgsinos
may be still be allowed in the framework of $\lambda^{''}$ type $R$-parity violating scenario. This scenario is quite
consistent with naturalness argument and is not possible in case of $R$-parity conserving or if $LLE$ and $LQD$ operators are included.
If the gluno decays to a sbottom and the mass difference between higgsinos and sbottom is less than the top quark ,
$ \tilde{b}_1  \rightarrow  b \tilde{\chi}_{1/2}$ will be dominant. In that case we will get 10 quarks from gluino pair
production and the bound is expected to be relaxed about 100 GeV. Additionally, in some cases, $\tilde{g} \rightarrow  g \tilde{\chi}_{1/2}$
branching can be significant ($\sim$ 10--20 \%) and it should be treated separately. Complications may arise, if
the higgsinos decay to top and bottom quarks. If a stop or sbottom squark is the LSP, the gluino will directly decay to a stop and sbottom
and the stop/sbottom will decay to 2 quarks. Existing study shows that if gluino branching to stop quark is
assumed to be unity,
gluino mass of 300 GeV and 250 GeV stop squark is still allowed by ATLAS same sign di-lepton search \cite{Han:2012cu}. \\

It is pointed out in ref. \cite{Han:2012cu} that the present search methods employed by CMS and ATLAS collaborations are not optimized
for $\lambda^{''}$ type $R$-parity violation and further improvement is possible. They suggest that for gluino decaying
to stop squark, the final state with single lepton with high jet multiplicity ($ N_{jet} \geq$ 7) and a large value of
$H_T$ variable can discover gluino mass up to 1 TeV at the present 8 TeV LHC run. It is also possible to
reconstruct stop squark in the gluino cascade. In the case where the gluino decays to $t\bar{t} \chi^0_1$ (and $\chi^0_1 \rightarrow q q q$) or $\tilde{t_1} \bar{t}$ (with $\tilde{t_1} \rightarrow  q q$),  final state consists of 18 or 10  partons respectively and even if the gluino is produced very close to rest, jet sub-structure
technique might be able to constrain the gluino mass\cite{Cohen:2012yc}. This method may improve gluino mass bound around
200 GeV assuming 5 fb$^{-1}$ data at 8 TeV. The analysis is carried out in ref. \cite{Curtin:2012rm} considers gluino decays to three light quarks and they show that color flow technique can be used for independent confirmation of ATLAS conventional search result.
They have shown that a gluino mass of up to 750 GeV can be probed at 8 TeV with 20 fb$^{-1}$ integrated luminosity using this method.
\\

\section{Grand Unification and $UDD$ $R$-parity Violation\label{sec: model}}

\begin{table}[t]
\begin{center}
{\renewcommand\arraystretch{1.15}
\begin{tabular}{|c|c|c|c|c|c|c|c|c|c|}
\hline
& $\mathbf {10}$& $\mathbf {5^*}$& $\Phi(\mathbf {5})$ & $\bar{\Phi}(\mathbf {5^*})$
& $\Sigma({\mathbf 1})$ & $U'({\mathbf 1})$ & $\bar U'({\mathbf 1})$ &$D'_i({\mathbf 1})$ &$\bar D'_i({\mathbf 1})$\\
\hline
$U(3)$ & ${\mathbf 1}$ & ${\mathbf 1}$
& ${\mathbf  {3^*}}$ & ${\mathbf  {3}}$ & ${\mathbf 8}+{\mathbf 1}$ & ${\mathbf  {3}}$ & ${\mathbf  {3^*}}$ & ${\mathbf  {3}}$& ${\mathbf  {3^*}}$\\
\hline
\end{tabular}
}
\end{center}
\caption{
The charge assignments of the model based on $SU(5)\times U(3)$.
We assign $U(1)$ charge $1/\sqrt{6}$ to the fundamental representation of $SU(3)$ except for $U'$ which has charge $-2/\sqrt{6}$.
}
\label{tab:GUT}
\end{table}%

In the previous sections, we showed that $UDD$ $R$-parity violation is consistent with current experimental constraints and then discussed naturalness in the context of different mass spectra. In this section, we will develop a model which justifies our consideration of only $UDD$ $R$-parity violation while neglecting other forms of $R$-parity violation. Our aim in this section is a GUT consistent model with the $UDD$ operator the only source of $R$-parity violation. Our jumping point is the $SU(5)\times U(3)$ product gauge unification model. We supplement this mdoel with additional quark like fields. These supplemental matter fields can be seen in Table \ref{tab:GUT}. We have not included the Higgs nor all of the matter needed to realize the doublet-triplet splitting. However, a more in depth discussion of the doublet-triplet splitting can be found in \cite{ProdUni} and is consistent with this section.
\subsection{$UDD$ $R$-parity Violation}
To generate $R$-parity violation, we have added three additional fields charged under the $U(3)$, $U'$, $D'_1$, and $D'_2$. If these fields are included, there are additional renormalizable interactions in the superpotential
\begin{eqnarray}
W=\lambda U'D'_1D'_2+\lambda \bar U'\bar D'_1\bar D'_2 +\lambda_{D_{1i}} {\bf 5}_i^*\Phi D'_1+\lambda_{D_{2i}}{\bf 5}_i^*\Phi D'_2\\ \nonumber  \quad\quad \quad +M_U\bar U' U' +M_{D_1}\bar D'_1D_1' +M_{D_2}\bar D'_2 D'_2
\end{eqnarray}
where ${\bf 5}^*_i$ contains the SM fields and $i$ is a family index.
As can be seen in the above superpotential, there are two $UDD$ like operators which are consistent with all the symmetries of the theory. $R$-parity violation is hidden in these operators until the gauge symmetries are broken to those of the SM\footnote{We assume the R parity of the hidden quarks, $U'$, $D'_{1,2}$, ${\bar U'}$ and ${\bar D'_{1,2}}$, are odd.}. In product unification, the vev of $\Phi$ breaks $SU(5)\times U(3)\to SU(3)_c\times SU(2)\times U(1)$ where $SU(3)_c$ is the diagonal subgroup of the $SU(3)$ subgroup of $SU(5)$ and the $SU(3)$ of the $U(3)$. The vev of $\Phi$ breaks the gauge symmetries to SM symmetries and generates mixing between the SM $D_i$ quarks and $D'_{1}$ and $D'_{2}$ in the form of a supersymmetric mass
\begin{eqnarray}
\lambda_{D_{1i}}v\bar D_iD'_{1,2}
\end{eqnarray}
where $v$ is the vev of $\Phi$. The SM $U$ quarks will also mix with the hidden sector quarks if we consider higher dimensional operators like\footnote{Since the GUT scale and Planck scale are relatively close and the need couplings of the $UDD$ operator are fairly small, this Planck suppressed operator is of the correct order of magnitude for our considerations.}.
\begin{eqnarray}
W=\frac{\lambda_{U'_i}}{M_{P}} {\bf 10}_i \bar \Phi\bar \Phi U=\frac{\lambda_{U'_i}}{M_{P}}v^2 \bar U_i U'
\end{eqnarray}
where $M_P=2.4\times 10^{18} {\rm GeV}$ is the reduced Planck mass and ${\bf 10}_i$ contains the SM fields. Again, the mixing of the $U'$ and the SM quarks is via the vev of $\Phi$. Because of this mixing, $U'$ and $D_{1,2}$ are not the mass eigenstates. Since they mix with the massless low scale quarks, they will each contain a component that corresponds to the light quarks. This component will then lead to $R$-parity violation in the low scale.

After $SU(5)\times U(3)$ is broken to the SM gauge groups, $U'$, $D'_1$, $D'_2$ are charged under $SU(3)_c\times U(1)$ and so contribute to the running of the SM gauge couplings. To avoid any problems with gauge coupling unification, we need the mass of the additional quarks to be close to or above the GUT scale. Furthermore, GUT scale masses will also assist us in suppressing the $R$-parity violating couplings as we will see below.

To determine the dependance of $R$-parity violation on the mass parameters, we give some details of the $U$ quark mass mixing
\begin{eqnarray}
W=M_{U'}\bar U' U' + \frac{\lambda_{U'_i}v^2}{M_{P}}\bar U_i U'=M_{U'} \bar U' U'+M_{U_i}\bar U_i U'.
\end{eqnarray}
Rotating these fields to their mass eigenstates, we find
\begin{eqnarray}
&&\bar U'\simeq \bar U_{m_0}-\frac{1}{M_U}\left(M_{U_1}\bar U_{m_1}+M_{U_2}\bar U_{m_2}+M_{U_3}\bar U_{m_3}\right)=\bar U_{m_0}-\frac{M_{U_i}\bar U_i}{M_U}\\
&&\bar U_{i}=\bar U_{m_i}+\frac{M_{U_i}}{M_u}\bar U_{m_0}
\end{eqnarray}
where $U_{m_0}$ is the mass eigenstate with non-zero mass and $U_{m_i}$ are the massless mass eigenstates. Analogous relations exist for  $D'_1$ and $D'_2$ found by substituting $\bar U_i\to \bar D_i$ and $\bar U'\to \bar D'_{1,2}$. The $R$-parity violating couplings in the low scale are then

\begin{eqnarray}
W=\frac{M_{U_i}M_{D_{1_j}}M_{D_{2_k}}}{M_U M_{D_1}M_{D_2}}\bar U_i \bar D_j \bar D_k
\end{eqnarray}
where $M_{D_{1_j}}=\lambda_{D_{1_j}}v$, $M_{D_{2_j}}=\lambda_{D_{2_j}}v$ and $M_{U_i}=\lambda_{U'_i}v^2/M_{P}$. By properly choosing the mass terms, a hierarchy among the couplings can be achieved. Using this hierarchy, we can safely select a single coupling that will be the dominant source of $R$-parity.

\subsection{Soft Masses of Product Gauge Unification}
 In the previous section, we showed how to generate the needed $UDD$ $R$-parity violating coupling using product gauge unification. In this section, we wish to give some discussion on possible mass spectra of this model. However, since we are interested in general features, we will not perform any parameter scans. The major difference of this model, from the typical approach to grand unification, is that the gaugino masses are not universal at the GUT scale\footnote{In the case with gaugino mass universality at the GUT scale, the low scale gaugino mass spectrum has the special relation
($M_1:M_2:M_3=1:2:6$) with a wino like lightest chargino and second lightest neutralino. These two particles
can decay to a bino like LSP by emitting a W or Z boson and thus have final states that contain $\slashed{E_T}$ or leptons. This
pushes up the mass bounds on the gluino to around 750 GeV for the CMS and ATLAS
7 TeV data  \cite{Asano:2012gj}.  }.  In fact, since the $SU(3)_c$ gaugino is a mixture of the $SU(3)$ subgroups of the $SU(5)$ and the $U(3)$, it can take on any value. The same goes for $U(1)_Y$ which is a diagonal subgroup of the $U(1)$ subgroup of $SU(5)$ and $U(3)$. Since the gauginos of $SU(3)_c$ and $U(1)_Y$ are arbitrarily deflected away from the unified gaugino mass scale, all three of the SM gauginos masses can have any value.

Because the gauginos are no longer unified, the typical hierarchy of the low-scale guagino masses found in other GUT models is not realized.  In fact, the gluino can even be taken as the LSP. This is crucial to our discussions of naturalness, because the stop masses will tend to be driven to the mass scale of the gluino. If the gluino is light, the large top Yukawa coupling will drive the stop quarks masses to smaller value.  A light stop will then lead to a more natural Higgs soft mass at the low-scale as well. On the other hand, if the gluino is heavy, the gluino mass term in the RG running will drive the stop masses heavy.  So the naturalness of a GUT SUSY is typically set by the gluino mass.

With the important parameters for naturalness being set by the gluino mass, all we need is a light gluino and the other parameters will flow to more natural values. But if one wishes to consider universal soft masses at the GUT scale, it may seem there is tension because the masses of the first and second generation squarks must be of order a TeV.  However, if the gluino is taken to be of order 500 GeV and the boundary soft masses are taken to be of order a TeV, the first and second generation squarks will remain of order a TeV because their Yukawa couplings are small and the stops will be driven light\footnote{Since the Yukawa coupling dependence of the left and right stop is different, one of the stops tends to be heavier. This difference can be offset to some degree by choosing the wino mass light and bino mass heavy. This hierarchy of gauginos will also tends to be necessary in order to generate radiative EWSB since a wino mass too large could prevent the Higgs soft mass from going negative.} because its Yukawa coupling is large. The Higgs soft mass will also tend to be smaller since the stops are lighter.

Now we wish to discuss the factors in generating the spectra discussed in section \ref{sec: pheno}. We will not discuss the details of the relative masses, but we will discuss where masses fall relative to the gluino mass.  Now, the stop can of course be driven lighter than the gluino because of its large Yukawa coupling. Depending on $\tan\beta$ and the wino and bino mass, it can be larger or smaller than the gluino at the weak scale even if its boundary mass is a TeV.  For large $\tan\beta$, the bottom Yukawa coupling is large and so the right handed sbottom can be made light.  A large bottom Yukawa coupling also aids in keeping the left and right stop masses equal, since the bottom Yukawa coupling contributes to the left stop running and not the right. The higgsino masses are set by the $\mu$ parameter.  The $\mu$ parameter is set by tuning it against the Higgs soft mass to get something of order $M_Z$.  So the higgsinos will tend to have masses of order the Higgs soft masses. Using the freedom of the wino and bino mass, a higgsino mass larger or smaller than the gluino can again be achieved,  a smaller value for $\mu$ being more natural, of course.

This model not only generates $UDD$ $R$-parity violation, but the hierarchy of the gauginos masses makes it possible to generate a natural low scale spectra while having soft mass universality.  This means we need not worry about the flavor problem of the soft masses.

\section{Conclusion}
Including the $R$-parity violating $UDD$ operator, we have discussed, on general ground, the more natural looking spectra allowed when this operator is included. Because $R$-parity violation removes the missing energy from the LHC signal and leads to large jet multiplicities, the constraints on the gluino tend to be much weaker. In fact, $UDD$ $R$-parity violation allows for many unique spectra that can have a gluino mass around 600 GeV as well as stops and higgsinos around 500 GeV. $UDD$ $R$-parity violating models are one of the last remaining hopes for natural SUSY.

To generate the $UDD$ operator without generating other $R$-parity violating couplings is non-trivial in the context of gauge coupling unification. In simple $SU(5)$ models the marginal $R$-parity violating operators are $\bar 5\bar 5 10$. In this context, it is impossible to generate $UDD$ without generating the operator $LLE$ and $LQD$. However, if the product group unification $SU(5)\times U(3)$ is considered, $SU(3)_c$ is the diagonal subgroup of the $U(3)$ and the $SU(3)$ coming from $SU(5)$.  This freedom allows us to have "$R$-parity violating couplings" in the $U(3)$ sector. Once $SU(5)\times U(3)\to SU(3)_c\times SU(2)\times U(1)_Y$, the $R$-parity violation is transmitted to the MSSM in the form of the $UDD$ operator alone.  Since gauge coupling unification is one of the important signatures of supersymmetry, product unification offers an interesting way to maintain grand unification and generate $R$-parity in the form of $UDD$. Other advantages of product group unification include non-universal gaugino masses. This is crucial since non-universal gaugino masses are necessary for generating more natural low scale mass spectra that are consistent with LHC constraints.

To this point we have neglected one important issue, the mass of the Higgs boson.  Since the stops are relatively light, the Higgs boson tends to be rather light.  However, if the Higgs boson is charged under an additional $U'(1)$ prime, a $125$ GeV Higgs can be realized. Furthermore, if the SUSY breaking mass of the $U'(1)$ gaugino is relatively small, the MSSM particles will be relatively unaffected by this additional $U'(1)$ and our analysis should still be valid. It would be interesting to see what other mechanism for enhancing the Higgs boson mass can be made compatible with the model we consider here. However, we have left the discussion of the Higgs mass to future work.

\section*{Acknowledgments}

S.M. and T.T.Y. were supported by the Grant-in-Aid for Scientific research from the Ministry of Education, Science, Sports, and Culture (MEXT), Japan (No. 23740169 for S.M. and No. 22244021 for S.M. \& T.T.Y.). This work was also supported by the World Premier International Research Center Initiative (WPI Initiative), MEXT, Japan. The work of J.L.E was supported in part by DOE grant DE--FG02--94ER--40823 at the University of Minnesota.


\end{document}